# DETERRENTS TO A THEORY OF QUANTUM GRAVITY

**Mario Rabinowitz**[1]

___________________________________________________________________________

As shown previously, quantum mechanics directly violates the weak equivalence principle in general and in all dimensions, and thus indirectly violates the strong equivalence principle in all dimensions. The present paper shows that quantum mechanics also directly violates the strong equivalence principle unless it is arbitrarily abetted in hindsight. Vital domains are shown to exist in which quantum gravity would be non-applicable. There are classical subtleties in which the strong equivalence principle appears to be violated, but is not. Neutron free fall interference experiments in a gravitational field reveal some evidence for a violation of the equivalence principle. Galileo's falling body assertion and the misconception it leads to, are examined,

___________________________________________________________________________

## 1. INTRODUCTION

Of the many different and assiduous approaches that have attempted to combine Einstein's general relativity (EGR) and quantum mechanics (QM) into a theory of quantum gravity (QG) none have been successful. This has not been for lack of due diligence, as some of the greatest minds in physics have worked on this problem for over three-fourths of a century since the groundbreaking work of Rosenfeld (1930). Stumbling blocks keep getting in the way of obtaining this holy grail of physics.

The strong equivalence principle (SEP) and the weak equivalence principle (WEP) each have clear meanings in classical mechanics (CM) and in QM. The WEP is unambiguous in both CM and QM. The WEP requires the independence or approximate independence of the orbiting mass m about the mass M. As previously shown (Rabinowitz, 2006 b) QM clearly violates the WEP, implying that QM violates the SEP. As we shall see, it is not possible to formulate the SEP in QM because QM violates the SEP both indirectly and directly. **Nevertheless the meaning of the SEP is comprehensive, and coincides in CM and QM that an accelerated frame can locally simulate an apparent gravitational field (and vice versa)**, even though there are violations, ambiguities and irrelevancies in QM.

[1] Armor Research, 715 Lakemead Way, Redwood City, California 94062-3922 USA
email: Mario715@gmail.com

The case of high gradient gravitational fields presents problems of non-applicability for both large size bodies in CM, and large width wave packets in QM when the width is larger than the region with approximately uniform field, as discussed in Sec. 3. In Sec. 2 we will find an interesting ambiguity if not an inconsistency related to the accelerated frame coordinate transformation in QM.

## 2. DIRECT VIOLATION OF THE STRONG EQUIVALENCE PRINCIPLE BY QUANTUM MECHANICS

Let us consider a simple transformation of coordinates to an accelerated frame for a particle of mass m. For a field-free region, the one dimensional Schroedinger equation is

$$\frac{-\hbar^2}{2m}\frac{\partial^2}{\partial x^2}\Psi(x,t) = i\hbar\frac{\partial}{\partial t}\Psi(x,t) \ . \tag{1}$$

To directly test the validity of the SEP in QM, we make a transformation to an accelerated reference frame of acceleration -g. The transformed coordinates are

$$x_a = x - vt - \tfrac{1}{2}gt^2 \text{ and } t = t_a \tag{2}$$

This transforms the wave function $\Psi(x,t)$ to $\Psi_a(x_a,t_a)$ such that

$$\Psi_a(x_a,t_a) = e^{iS}\Psi[x(x_a,t_a),t(x_a,t_a)], \tag{3}$$

where $e^{iS}$ is introduced to represent a possible shift in phase. $\Psi(x,t)$ does not contain the product $mg$ nor higher order terms in g in any term since they are not present in eqs. (1) and (2). Therefore the only way $\Psi_a(x_a,t_a)$ could contain $mg$ and higher order terms in g is for them to be arbitrarily introduced in hindsight into the change in phase factor $S$. Without an artificial introduction of $mg$, $\Psi_a(x_a,t_a)$ also represents a field-free solution, since a gravity potential energy term of the form $mgx_a$ could not otherwise be present in the Schroedinger equation. For compliance with the SEP, eq. (3) would have to result from a Schroedinger equation of the form

$$\frac{-\hbar^2}{2m}\frac{\partial^2}{\partial x_a^2}\Psi_a(x_a,t_a) + mgx_a\Psi_a(x_a,t_a) = i\hbar\frac{\partial}{\partial t}\Psi_a(x_a,t_a) \ . \tag{4}$$

But this is not directly possible because the transformed wave function $\Psi_a(x_a,t_a)$ does not contain the product $mg$ and higher order terms in g until they are introduced indirectly by hand into $S$. The SEP can only be obeyed (Colella, Overhauser, and Werner, 1975), in hindsight by phenomenological curve fitting to the experiment in subjectively setting

$$S = (mgx_a t/\hbar) + (mg^2 t^3/6\hbar). \tag{5}$$

Without knowing the required answer in advance, S could be tailored so that the wave function results from a Hamiltonian with a different potential energy. There is even a limit to what can be put in by hand since a uniform acceleration transformation of reference frame can only be represented by a uniform gravitational field, i.e. a linear gravitational potential energy in the Hamiltonian of any of the quantum differential equations. This limited degree of arbitrariness or malleability in QM allows it to give the correct answer on an *ex post facto* basis, and not directly from the reference frame transformation.

A similar transformation of reference frame argument that QM directly violates the SEP can be made for the Dirac and the Klein-Gordon equations. Or one can infer that since both the Dirac and the Klein-Gordon equations reduce to the Schroedinger equation, they also directly violate the SEP in the same way that the Schroedinger equation does. Unequivocally, QM directly violates the WEP both non-relativistically and relativistically (Rabinowitz, 2006b). This goes further than just motion in a gravitational field as the Schroedinger, Dirac and Klein-Gordon equations have mass dependent field-free motion in strong conflict to both EGR and Newtonian mechanics where field-free motion is mass independent. Non-relativistically, the wave function that satisfies eq.(1) for a free particle is

$$\Psi(x,t,m) = \frac{1}{\sqrt{2\pi}} \int_{-\infty}^{\infty} [\Phi(k)]_{t=0} \exp\left[i\left(kx - \frac{\hbar k^2}{2m}t\right)\right]dk, \text{ where} \tag{6}$$

$$\Phi(k) = \frac{1}{\sqrt{2\pi}} \int_{-\infty}^{\infty} e^{-ikx}\Psi(x,m)dx, \tag{7}$$

and $k = p/\hbar$ is the wave vector.

In 1930 Schroedinger noted that Zitterbewegung results from wave packet solutions of the Dirac equation for relativistic electrons in free space due to an interference between positive and negative energy states. Zitterbewegung, a helical motion of spin 1/2 particles, violates the SEP. For an electron, it is a fluctuation (at the speed of light) of the position around its uniform motion, with a mass dependent angular frequency of $2mc^2/\hbar \sim 10^{21}$ Hz.

A direct non-compliance by QM of the SEP is in accord with the direct violation of the WEP by QM. Synthetically making QM comply with the SEP, results in the anomalous quandary that although the SEP implies the WEP classically, it appears not to imply the WEP quantum mechanically rather than being judged to not comply with QM itself. Since the SEP is the cornerstone of Einstein's general relativity, and it is not unequivocally clear that QM honors the SEP, quantum mechanics may not be compatible with EGR and a theory of quantum gravity may not be possible.

### 3. Vital Domains in Which Quantum Gravity Would Not be Applicable

In addition to the violation of the SEP, there is another serious problem to the development of a theory of quantum gravity. It is the non-applicability of the SEP in QM. Quantum gravity is vitally needed for a proper theory of little black holes (LBH) and their radiation (Rabinowitz, 2001, 2006b). For problems where QG is necessary there may not be a region of space containing an approximately uniform field even comparable in size to the body's QM wave packet to permit the application of the SEP. Such is the case for very large gradients of the gravitational field as when little black holes (LBH) decay due to their radiation (Rabinowitz, 2006a). The wave packet of width $\Delta x$ should be smaller than the smallest region in which the gravitational field $\Im$ is sufficiently uniform in an inertial frame to approximate the field by an accelerated frame. In non-relativistic terms we require:

$$\Delta x \equiv x_2 - x_1 < [\Im(x_2,y,z,t)/\nabla\Im] - [\Im(x_1,y,z,t)/\nabla\Im] \tag{8}$$

In the simplest case for a particle of momentum p, the dispersion of the position coordinate $x_i$ (the particle's position variance as a function of time) for a Guassian wave packet, is of the form

$$\Delta x_i = \Delta x_{i0}\left[1 + \frac{a^2 t^2}{(\Delta x_{i0})^4}\right]^{1/2}, \tag{9}$$

where $a = \left(\frac{\partial^2 w}{\partial k^2}\right)_{k=k_0}$, $w$ is the angular frequency of the particle wave, and $k = p/\hbar$ is its wave vector. [The possibility should be kept in mind that for large enough mass/mass density self-gravitation can impede or in some cases even stop the spreading of the wave packet.] In the parlance of Einstein's general relativity (EGR), a small region of curved space-time must exist that appears Euclidean (flat) within the approximation validity in QM that the region is > the wave packet width $\Delta x$ for the accuracy of the calculation being made. Otherwise, the conventional concept of QG is not applicable, rather than wrong.

This is also the case for orbits of gravitationally bound atoms (Rabinowitz, 2006a) made of LBH or other highly compact matter, due to large gravitational gradients and when the width $\Delta x$ (e.g. standard deviation of a Guassian distribution) of the wave packet is comparable to the circumference of the orbit. The problem may go deeper than the small number of wavelengths in an inner gravitational orbit. Taking the limit of $\Delta x \rightarrow \sim 0$ may bring in not only QM inconsistencies due to the uncertainty principle, but even corresponding classical inconsistencies related to self-energy.

There may be a similar problem for such gravitational orbits with respect to the expectation values not corresponding to their classical counterparts contrary to Ehrenfest's theorem. The concept of a classical trajectory does not apply in general. Even in the simplest case of free fall in a uniform gravitational field, the QM mathematical complexity of achieving correspondence between the expectation values and the easily obtained classical values is formidable. It appears not to be achievable for highly non-uniform gravitational fields, even without regard to the question of whether or not the SEP is applicable. Expectation values in QM result from integrals over a hypersurface in all space for a given value of time – and their compatibility with the local notions of EGR are questionable. Furthermore as we shall see in Sec. 5, there is no evidence that the SEP is experimentally obeyed in the realm most relevant to QG – that of large fields with large gradients. Such QM fulfillment may not be incontrovertible as the concept of free fall observation in the quantum domain is not well-defined even for weak fields, with observation (measurement) interfering with the free fall.

## 4. Relevance of Semi-Classical Mechanics

To illustrate the violation of the weak equivalence principle (WEP) in three and higher dimensions, I utilized semi-classical mechanics to calculate the quantized orbits of gravitationally bound atoms (Rabinowitz, 2006b). Although this avoided undue complexity, two independent questions may be raised about the validity of doing this. One relates to the use of semi-classical mechanics in this domain since to allow for atomic size orbits ($\in 10^{-10}$ m) one must utilize even smaller highly compact dense bodies such as little black holes (LBH). The other relates to the very validity of semi-classical mechanics itself.

With respect to the first question, it seems reasonable to challenge the use of semi-classical physics at such a small scale and high energies. However, as measured at large distances, the gravitational red shift substantially reduces the impact of high energies near LBH. Argyres et al (1998) argue that "...one can describe black holes by semi-classical physics down to much smaller masses of order of the fundamental Planck scale [$10^{-35}$ m]... ,"

With respect to the second question, it is true that semi-classical mechanics breaks down for chaotic (non-integrable) systems. In simple terms the semi-classical Bohr-Sommerfeld condition $\oint p \cdot dl = jh$ breaks down for non-periodic orbits, where $p$ is the canonical momentum and j is an integer. This was first shown by Einstein (1917) and rediscovered by J.B. Keller (1958). This breakdown does not invalidate semi-classical mechanics, but rather restricts its domain of validity. For example, as shown by Sommerfeld this theory does amazingly well in predicting both the fine and hyperfine structure of the energy levels of the hydrogen atom. In this case, its precision is comparable to that given by the Dirac equation. Thus in my semi-classical calculation of periodic atomic size LBH gravitational orbits (Rabinowitz, 2006 a, b) the same mass dependency and basically the same results are obtained as from the Schroedinger equation. It has been previously noted that the Bohr-Sommerfeld condition violates the WEP much the same as does the Schroedinger equation since

$$\oint p \cdot dl = \oint (mv) \cdot dl = jh \Rightarrow \oint v \cdot dl = j\frac{h}{m}, \tag{10}$$

where $v = \partial H/\partial p$ is the canonical velocity. (In Sec. 6, a rotational frame yields $v = dr/dt + \mathbf{w} \times dr/dt$, which is analogous to $v = dr/dt + (e/m)A$ for the canonical velocity of a particle of charge e and mass m in a magnetic field of vector potential $A$.) Since eq. (10) is the starting point for an n-dimensional analysis, the variables of periodic orbits are quantized in terms of both m and Planck's constant $h$ in violation of the WEP in all dimensions.

## 5. Classical Subtleties

There are even classical subtleties in which the SEP appears to be violated, but is complied with due to deeper sagacious conceptualization. The SEP implies that a charged body will free-fall accelerate in a uniform field at the same rate as a neutral body. Yet the charged body must lose energy while falling because it is radiating. Hence at first blush one expects it to fall more slowly than the neutral body which is not encumbered by the need to radiate. Rohrlich (1965) argues that in compliance with the SEP, the two bodies fall at the same rate due to the needed extra energy coming from the gravitational distortion of the charged particle's Coulomb field. One may raise the question: What about the radiation reaction force? Remarkably, EGR yields a zero radiation reaction force on the charged body for free-fall in a static, uniform gravitational field.

There are other classical subtleties in which the SEP would definitely be violated, but they have not borne out to date. One such possibility was called "The Fifth Force." Fischbach et al (1986) in their analysis of the Eötvös (1892) experiment, that gravitational attraction is independent of the composition of attracting bodies, thought they found a small error in his work related to the baryon number (number of protons and neutrons) of the tested material. Although this would only be a small violation of the SEP, it would nonetheless have been a violation. If there were such a violation, there might have been a band-aid fix that the SEP is still valid, but that one needs to superimpose an additional fifth force on gravity that operates in the range of ~ 100 meters. Until that time gravity was not well-tested in this range, and this relatively localized force would only slightly affect EGR. However it might have a major effect on quantum gravity. This dilemma did not materialize as new experiments did not support the hypothesis of a fifth force.

Most bodies fall at the same rate on earth, relative to the earth, because the earth's mass M is extremely large compared with the mass m of most falling bodies for the reduced mass $m = mM/(m+M) \approx m$ for $M >> m$. The body and the earth each fall toward their common center of mass, which for most cases is approximately the same as relative to the earth. As discussed in (Rabinowitz, 1990), in principle the results of a free fall experiment depend on whether falling masses originate on earth, are extraterrestrial, are sequential or concurrent, or are simultaneous for coincident or separated bodies, etc. When falling bodies originate from the earth, all bodies (light and heavy) fall at the same rate relative to the earth in agreement with Galileo's view because the sum m + M remains constant. When extraterrestrial bodies fall on earth, heavier bodies fall faster relative to the earth making Aristotle correct and Galileo incorrect. The relative velocity between the two bodies is $v_{rel} = [2G(m+M)(r_2^{-1} - r_1^{-1})]^{1/2}$, where $r_1$ is their initial separation, and $r_2$ is their separation when they are closer.

Even though Galileo's argument (Rabinowitz, 1990) was spurious and his assertion fallacious in principle -- that all bodies will fall at the same rate with respect to the earth in a medium devoid of resistance, it helped make a significant advance in understanding the motion of bodies. Although his assertion is an excellent approximation that can be exactly true in some cases (as explained in the preceding paragraph), it is not true in general. Galileo's alluring assertion that free fall depends solely and purely on the milieu and is entirely independent of the properties of the falling body, led Einstein to the geometric concept of gravity. Nevertheless in EGR, m can alter the ambient field because m is also a source (as is its field) of a gravitational field – albeit negligible for m << M. Due to the non-linear character of EGR, the effect is much bigger than the superposition of the fields of m and M. Quantum mechanics has its subtleties such as motion through classically forbidden regions that may prevent a purely geometrical construct for gravity. [In tunneling there is a possible superluminal velocity (cf. Sec. 7), the kinetic energy < 0, and the momentum is imaginary.] At the very least, one has to be careful in choosing reference frames that incorporate quantum corporeal objects.

## 6. Neutron Free-Fall Interferometry Experiments

Neutron interferometry, gravity induced, quantum interference experiments were used to test the SEP in QM. An almost mono-energetic beam of thermal neutrons is split into two parts, by nearly perfect silicon crystals so that each part traverses a path of slightly different gravitational potential. The concept of a classical trajectory applies because the quantum wave packet is smaller than the loop formed by the two trajectories (cf. my comments at the end of Sec. 8 for a large wave packet). The first experiment to show that gravity affects the quantum mechanical neutron phase was the beautiful COW experiment reported in 1975 using neutron interferometry (Colella, Overhauser, and Werner, 1975), but it neither tested nor verified the SEP. The neutrons fell in the earth's gravitational field, but in this particular test no accelerated frame experiment was purposely conducted. They do make the statement, "this experiment provides the first verification of the principle of equivalence in the quantum limit." However, this is only a theoretical inference since no accelerated frame experiment was intentionally done, although the COW experiment inadvertently coupled the interferometer to the earth's rotation as a small secondary effect.

The phase difference $\Phi$ between an upper neutron path and a lower neutron path in the interferometer is

$$\Phi_{up} - \Phi_{low} = \frac{-m_i m_g g d' \bar{\lambda} \ell \sin f}{\hbar^2} = \frac{-m^2 g d' \bar{\lambda} \ell \sin f}{\hbar^2}, \quad (11)$$

where m is the neutron mass, g is the local gravitational acceleration, $d'$ is the horizontal neutron path, $\bar{\lambda} \sim 1 \overset{o}{A}$ is the neutron reduced de Broglie wavelength, and $\ell \sin f$ is the vertical part of the neutron path. The angle $f$ [varied from 0 to $\pi/2$, and from 0 to $-\pi/2$] determines the frequency of oscillation of the neutron interference pattern, and is the angle by which the interferometer is rotated about the incident beam direction. Gravity, rotation, and inadvertent bending of the silicon crystal all contribute to the oscillation frequency. It should be noted that the phase shift depends on *m*, as it does in all quantum gravitational effects, in direct violation of the WEP, and indirect violation of the SEP.

The first experiment to attempt an assessment of the SEP was another elegant neutron interferometry experiment, using the earth's axial rotation as the accelerated frame (Staudenmann et al, 1980). The interferometer was limited by bending (warping) due to its own weight, and would be similarly limited by acceleration. It is not clear whether or not such experiments strictly test the SEP because the SEP applies to linear acceleration and does not rigorously apply in spinning reference frames. Furthermore a small effect was observed that relied upon a $10^9$ amplification. For their thermal neutrons, the effect of the earth's rotation on the neutron phase was about 2.5% of the earth's gravitational effect. Even this small effect is surprisingly large since the effect of the earth's rotation on the neutron trajectory is only $\sim 10^{-9}$ the displacement due to gravity as we shall see.

Neglecting negligible corrections (relativistic, non-uniformity of the earth's gravitational field, etc.), the Hamiltonian for neutrons of inertial mass $m_i$ and gravitational mass $m_g$ is

$$H = \frac{p^2}{2m_i} + m_g g_0 x - \mathbf{w} \cdot (r \times p) , \qquad (12)$$

where $p = m_i \frac{dr}{dt} + m_i \mathbf{w} \times r$ is the canonical momentum, $(r \times p)$ is the angular momentum of the neutron with respect to the center of the earth, $g_0 = 9.80$ m/sec$^2$ is the local gravity acceleration, and $\mathbf{w}$ is the angular spin velocity of the earth. From the Hamiltonian of eq. (12), we obtain the classical equation of motion by combining the results of Hamilton's equations, $v = \partial H / \partial p$ and $dp/dt = \partial H / \partial r$ for a neutron in the rotating frame of the earth:

$$m_i \frac{d^2 r}{dt^2} = -m_g g_0 \hat{x} - m_i \mathbf{w} \times (\mathbf{w} \times r) - 2 m_i \mathbf{w} \times \frac{dr}{dt}, \qquad (13)$$

where the last term represents the Coriolis force giving the apparent deflection of a moving neutron in a rotating frame, and the penultimate term represents the centrifugal force acting on the neutron in this rotating frame. Because of the earth's low angular velocity, the linear Coriolis term dominates over the quadratic centrifugal term. The Coriolis acceleration $2\mathbf{w} \times dr/dt \sim$ $\sim 10^{-1} m/\sec^2 \sim 10^{-2} g_o$, and the centrifugal acceleration $\mathbf{w} \times (\mathbf{w} \times r) \sim 10^{-2} m/\sec^2 \sim 10^{-3} g_o$. These accelerations are $<< g_o$, and miniscule in terms of the realm of QG.

Solving eq. (13) for the radial position of a neutron

$$r \approx r_0 + v_0 t - \tfrac{1}{2} g t^2 + \tfrac{1}{3} \mathbf{w} \times g t^3. \tag{14}$$

Their experiment assumed that there is an effective gravitational acceleration g which is relevant to the SEP, due to the earth's gravity and spin rotation where

$$g = g_0 \hat{x} + \frac{m_i}{m_g} \mathbf{w} \times (\mathbf{w} \times r) + \frac{2 m_i}{m_g} \mathbf{w} \times \frac{dr}{dt}. \tag{15}$$

From eq. (13) the ratio of the neutron displacement due to gravity and angular motion is

$$\frac{\tfrac{1}{3} \mathbf{w} \times g t^3}{\tfrac{1}{2} g t^2} \sim 10^{-9}, \tag{16}$$

where $\mathbf{w} = 2\pi/[(23 hr 56 min)(3600 sec/hr)] = 7.29 \times 10^{-5}$ /sec, and $t \sim 5 \times 10^{-5}$ sec for thermal neutron transit time through the interferometer. It is remarkable that such a small relative displacement $\sim 10^{-9}$ due to Coriolis and centrifugal acceleration translates into as much as a 2.5% effect on the quantum mechanical neutron phase. Let us next check the tiny effects of EGR.

In EGR there is no length change perpendicular to a gravitational field; and a length contraction parallel to the field. (This is analogous to length contraction in a moving frame for special relativity.) The relative change in length of the neutron trajectory in the direction of the earth's gravitational field is

$$\frac{\Delta d}{d} \approx \left( \frac{GM}{c^2 (R+d)} \right) - \left( \frac{GM}{c^2 R} \right) \approx \frac{-GMd}{c^2 R^2} = \frac{g_o d}{c^2} \sim -10^{-24}, \tag{17}$$

where the universal gravitational constant $G = 6.67 \times 10^{-11}$ n-m$^2$/kg$^2$, the mass of the earth $M = 6.0 \times 10^{24}$ kg, the earth's radius $R = 6.4 \times 10^6$ m, the speed of light $c = 3 \times 10^8$ m/sec, and the length of the neutron trajectory in the earth's field $\ell \sim 10^{-8}$ m. This is small because the earth's field is small.

Time runs slower in a frame of higher gravitational potential relative to a lower potential. (This is analogous to time dilation in a moving frame for special relativity.) The EGR relative time change due to a change in the earth's potential for neutron's at a height difference $d$ is equally small:

$$\frac{\Delta t}{t} \approx \frac{g_0 d}{c^2} \sim 10^{-24}. \tag{18}$$

Although the EGR calculations are not shown (Greenberger and Overhauser, 1980) so that one has no idea of the smallness of the effect, the statement is made: "The entire interference effect in the COW experiment can be attributed to the difference between the time on a clock moving along with one beam and the time on a clock moving along with the other beam." It is remarkable that this minuscule time difference between the times in the reference frame of the two interfering neutron beams could account for the interference pattern. It is amazing that it should be able to do so without recourse to a quantum mechanical explanation.

A fly entered the ointment as a result of the much more precise gravitational neutron interferometry experiment of (Littrell, Allman, and Werner, 1997). The two previously referenced experiments had an uncertainty of about 1% and seemed to be in accord with the equivalence principle. The statistical errors together with the estimated and measured uncertainties in this experiment are about an order of magnitude lower at 0.1%. They observed a discrepancy between the theoretically predicted and experimentally measured values of the phase shift due to gravity at the 1% level.

They felt that "the most likely explanation for this discrepancy is that the way in which the effects of strains and bending of the interferometer have been modeled is too naive." For them: "Another, less likely source for the discrepancy is a difference in the way in which centrifugal force acts in classical and quantum mechanics." Their experiment may indicate a tiny quantum mechanical difference between inertial mass $m_i$ and gravitational mass $m_g$. Such a difference may manifest itself in the phase difference given by eq. (11). If this is the case, it would be an experimental violation of the WEP in QM, indirectly implying a violation of the SEP. An experimental violation of the SEP in QM (albeit indirect) would be a further indication of the incompatibility of QM with EGR because the SEP is the cornerstone of EGR.

## 7. Discussion

In Sec. 2 of this paper, we found that the SEP is directly violated by QM since there is no SEP in QM, unless it is put in by hand. There is even a limit to what can be put in by hand as explained in Sec. 2. In my preceding paper (Rabinowitz, 2006b), we found that the WEP is directly violated by QM and this indirectly implies violation of the SEP. This was done with the understanding that if there is an SEP in QM, then it would imply the WEP. But since the WEP is violated in QM, the SEP must also be violated, i.e. there is no SEP in QM without a kind of phenomenological fitting or adjustment of variables. The results of this paper and my preceding paper are thus consistent and in accord.

There are two possible conclusions from the fact that QM violates the WEP. One is that somehow there is no need in QM for the SEP to imply the WEP, no matter how reasonable this implication is classically. The other is that in QM the SEP should imply, but does not imply the WEP because the violation of the WEP in QM is a clear indicator that QM violates the SEP. This latter conclusion is by far the more consistent. The violation of the WEP in QM also implies violation of the equality of inertial and gravitational mass.

Newtonian gravity (NG) has a possibly irreconcilable disparity (Rabinowitz, 2005) with EGR, and similarly for the Schroedinger equation of non-relativistic QM (NRQM) with respect to relativistic QM (RQM) as embodied in the Dirac and Klein-Gordon equations. The speed of gravity is unreasonably infinite in both NG and in NRQM. Because NG works so well, EGR must or should reduce to NG in the weak-field and low-velocity limits. Similarly Because NRQM works so well, RQM must or should reduce to NRQM in the low-speed limit $v \ll c$. Yet it is not clear that gravity can or should go from the speed of light, c, to infinite speed in these limits. As previously discussed (Rabinowitz, 2005), there are situations where speeds $v \gg c$ are found in EGR and in QM, which may be pointing to internal inconsistencies. Furthermore, conclusive experiments show that photons can tunnel through a barrier with a group velocity that greatly exceeds the speed of light in vacuum (Chiao and Steinberg, 1997). In one experiment, a photon traveled with superluminal speed through a barrier as compared with a control photon that went the same distance in vacuum. Similar results were obtained by other groups using microwaves, and femtosecond lasers. In one experiment an intelligible microwave version of Mozart's 40th symphony tunneled through a barrier at almost five times the speed of light. Since cause and effect were not inverted (e.g. the symphony did not play backwards), these experiments may indicate invalidity of the proofs that a signal velocity greater than c would invert causality.

If it is necessary to abandon the space-time continuum to avoid unmanageable infinities that plague attempts at a theory of QG, then jumps from one discontinuous point to another may well occur at superluminal speeds. Perhaps it is time to consider possible exceptions to the universal speed limit c in the development of a theory of QG. This is especially so at very small distances such as at the Planck length, $< \sim 10^{-35}$ m.

## 8. Conclusion

Of the many problems that are troublesome deterrents to a theory of quantum gravity, it appears to me that the fundamental problem lies in the reconciliation of quantum mechanics with the strong equivalence principle. Attempts to derive a theory of quantum gravity that are not based upon the SEP (or a reasonable facsimile) are self-contradictory. Theoretical compatibility of the present form of QM with the SEP is questionable. The WEP is always clearly violated by QM. The SEP can be saved in QM by benefit of hindsight when arbitrary abetting terms are put into the wave function by hand. The procedure is like phenomenological curve-fitting, and is not done directly from the reference frame transformation. It is unlikely that it can be done without *ex post facto* input.

This *ex post facto* input leads to a yet deeper problem. In QM the transformation to an accelerated reference frame is indeed explicitly independent of the accelerated mass leading to an explicit violation of the SEP, since the transformed wave function corresponds to the solution of a Hamiltonian without a gravitational potential. Hindsight calculation of the phase factor implicitly brings this transformation into compliance with the SEP, allowing the SEP to exist in QM. Yet QM explicitly violates the WEP with a mass dependence for gravitational free-fall, thus implicitly violating the SEP, and saying that the SEP does not exist in QM. If the hindsight calculation were proper, both SEP and not-SEP could be proven to exist in QM at the same time. Such a basic logic contradiction would imply a fundamental inconsistency that goes further than either a wave-particle duality or an incompatibility with EGR. It would point to an internal inconsistency in QM. Furthermore, it has not been established (and probably cannot be established) by experiment that quantum mechanics complies with the SEP for a large gravitational field with a large gradient.

As shown in Sec. 3, vital domains exist in which quantum gravity would be non-applicable, rather than violating the SEP. Whether a theory of quantum gravity were to be founded upon the SEP or some other principle, there are relevant questions that QG could not answer regardless of what it is founded upon. This is because of the very nature of QM that is fundamentally non-local.

As discussed in Sec. 6, the 1997 neutron interferometry experiments seem not to support the SEP for even an extremely weak gravitational field. This is far from the case of interest in QG. If instead of thermal neutrons they could have used slow cold neutrons, the quantum wavelength of the neutrons and their wave packets would have been too large to consider completely separate trajectories in the interferometer. This would exclude an explanation purely in terms of EGR, and quantum effects would be expected to result in a clear experimental violation of the SEP in QM. The inevitable ramification of the strong equivalence principle is that gravity is exclusively due to the geometry of space-time curvature, but this appears not to be the case at the quantum level.